\documentclass[12pt,preprint]{aastex}
\usepackage{graphics}
\usepackage{epsfig}

\shorttitle{The Distance and Age of Kes 73/AXP 1E 1841-045}
\shortauthors{Leahy \& Tian}

\begin{document}

\title{The Distance and Age of the SNR Kes 73 and AXP 1E 1841-045} 

%\email{tww@iras.ucalgary.ca}
\author{W.W. Tian\altaffilmark{1,2}, D.A. Leahy\altaffilmark{2}}
\altaffiltext{1}{National Astronomical Observatories, CAS, Beijing 100012, China, tww@iras.ucalgary.ca}
\altaffiltext{2}{Department of Physics \& Astronomy, University of Calgary, Calgary, Alberta T2N 1N4, Canada} 
 
\begin{abstract}
We provide a new distance estimate to the supernova remnant (SNR) Kes 73 and its
associated anomalous X-ray pulsar (AXP) 1E 1841-045. 
21 cm HI images and HI absorption/ emission spectra from new VLA observations, and $^{13}$CO 
emission spectra of Kes 73 and two adjacent compact HII regions (G27.276+0.148 and
 G27.491+0.189) are analyzed. 
The HI images show prominent absorption features associated with Kes 73 and the 
 HII regions. The absorption appears up to the 
tangent point velocity giving a lower distance limit to Kes 73 of 7.5 kpc, which has 
previously been given as the upper limit. Also,
G27.276+0.148 and G27.491+0.189 are at the far kinematic distances 
of their radio recombination line velocities.
There is prominent HI emission in the range 80--90 km/s for all three objects. 
The two HII regions show HI absorption at $\sim$ 84 km/s, but there is no absorption
in the Kes 73 absorption spectrum. This implies an upper distance limit of $\sim$9.8 kpc 
to Kes 73. This corrected larger distance to Kes 73/ AXP 1E 1841-045 system leads to a
refined age of the SNR of 500 to 1000 yr, and a $\sim 50\%$ larger AXP X-ray luminosity. 
  
\end{abstract}


\section{Background and Data}
Because H I atomic clouds are broadly distributed throughout the
Galactic plane (Dickey \& Lockman 1990), 21 cm H I
observations can directly provide distance constraints to
Galactic plane objects based on an axisymmetric rotation
curve model of the Galaxy. However, such a kinematic model causes the near-far kinematic distance ambiguity in the inner Galaxy, i.e. each radial velocity along a given line-of-sight corresponds to two distances equally spaced on either side of the tangent point. For most Galactic 
 HII regions where recombination lines are detected, the ambiguity is easily solved by comparison of the radial velocity of the recombination line with the highest velocity of the HI absorption in the HII region's direction, because cold HI atomic clouds in front of an HII region absorb the broad-band thermal bremsstrahlung radio continuum emission from the background HII region (Kolpak et al. 2003). For other objects of spectral-line radiation, CO observations have helped to reduce the ambiguity, because a CO molecular cloud behind an object will not produce HI absorption feature in the object's absorption spectrum. As a widely used tool to measure distance, 21 cm HI observations can generally provide a distance estiamte to precision of a few hundred pc. The circular rotation curves model for estimating distances is not always accurate, because any individual objects have its own peculiar velocity besides rotation (i.e. random cloud-to-cloud motions in the interstellar matter and systematic stream motions along spiral arms, Brand \& Blitz 1993), e.g. noncircular motions can put relatively nearby emission at negative velocities, and cause absorption velocity larger than the tangent-point velocity (McClure-Griffiths \& Gaensler 2005)  

Another challenge to the method is to construct a real HI absorption spectra to a Galactic object. The methods to construct HI absorption spectra of a compact Galactic source are discussed by Dickey and Lockman (1990). For a general case, i.e. including extended sources, the HI absorption spectrum can be found by the formula (Tian et al. 2007b; Leahy \& Tian 2008 for detail):
$\Delta T$ = $T^{HI}_{off}$-$T^{HI}_{on}$ = ($T^{c}_{s}$-$T^{c}_{bg}$)(1-$e^{-\tau}$).\\ $T^{HI}_{on}$ and $T^{HI}_{off}$ are the average brightness temperature of many spectra from a selected area on a strong continuum emission region of the target source and an adjacent background region (i.e. excluding the strong continuum emission area).  $T^{c}_{s}$ and $T^{c}_{bg}$ are the average continuum brightness temperatures for the same regions respectively. $\tau$ is the optical depth from the continuum source to the observer along the line-of-sight. When there is no continuum emission in the background region, this reduces to the standard formula (for continuum subtracted maps).
Spatial variation in the background HI emission over the target source region (due to occasionally patchy distribution, or unsolved emission structure) may cause spurious absorption features, and produce significant uncertainty in the measurement of the absorption spectrum. Faint continuum emission at 1.4 GHz from the target source can also cause too large noise to construct a reliable absorption spectrum. However, the high resolution observations from interferometers (e.g. Very Large Array (VLA) and Very Long Baseline Array (VLBA)) can reduce the uncertainty and produce high sensitivity absorption spectra (Dickey \& Lockman 1990). There still exists possibility of significant uncertainties in the absorption spectrum due to the
differences in distribution of very small HI clouds along the line of sight-to-source and background regions even using interferometer's observations. For HI maps from interferometer data, one has a nearly unlimited choice of source regions and background regions, however it makes it most sense to minimize the differences in HI distribution between the source region and the background regions by choosing them to be adjacent. In any case, CO observations, however, can help to reduce and explain the uncertainties of the HI absorption spectra, because CO emission spectra can show different emission features along either source or background line-of-sight (especially sensitive to small high brightness-temperature CO clouds).

Kes 73 is a very young and small (4 arcmin diameter) shell-type radio supernova remnant 
(SNR) with clumpy X-ray emission and a compact central source (Helfand et al. 1994).
The central source of Kes 73,  1E 1841-045, was discovered to be an anomalous X-ray 
pulsar (AXP) by its 11.8 second pulsations and high spin-down rate (Vasisht \& Gotthelf 1997)
and young characteristic age, $\sim$ 4.7 kyr. 
Further observations confirmed that it is an AXP with a superstrong magnetic field 
($\sim 7 \times 10^{14}$ G, Gotthelf $\&$ Vasisht 1999). AXPs and soft gamma-ray
repeaters are interpreted as magnetars (Thompson \& Duncan 1996). 
For an alternate explanation of the AXP and SGR phenomena involving accretion of 
a fall-back crust onto a quark star see Ouyed et al. (2007), for the case of a 
magnetically supported shell, and Ouyed et al (2006), for the case of a rotationally 
supported torus.

Based on the analysis of the X-ray spectrum of Kes 73, Gotthelf $\&$ Vasisht (1997) inferred 
an age for the SNR  of $<$ 2.2 kyr which is less than half the characteristic age
 of AXP 1E 1841-045: this emphasizes the inaccuracy of characteristic age as a measure of
true age.  
The distance to Kes 73 was determined by Sanbonmatsu \& Helfand (1992) to be
in the range of 6 -- 7.5 kpc by means of a 21 cm HI absorption measurement. 
They employed a commonly used technique to determine distance, by comparing an absorption
 spectrum toward Kes 73  with absorption seen toward a nearby bright source. 
However, this technique must be used with care, since there may be significant differences 
between the distributions of HI along the line-of-sight to the target source and 
the comparison source. In this case, multiple comparison lines-of-sight can determine
the correct distance.

In this paper, we revise the distance to Kes 73 based on the HI absorption and emission 
spectra and $^{13}$CO spectra of Kes 73 and nearby sources. The data that we use
come from 
the 1420 MHz continuum and HI-line observations of the VLA Galactic Plane Survey 
(Stil et al. 2006; Tian et al. 2007a) and from the $^{13}$CO-line (J = 1-0) observations of the Galactic 
ring survey (Jackson et al. 2006). 
The VLA Galactic Plane Survey data was taken with the VLA in D array configuration,
with synthesized beamwidth of 60$^{\prime\prime}$, spectral resolution of 1.56 km/s, channel width of 0.824 km s$^{-1}$,  and the noise per velocity channel of 2 K. Total bandwidth of the observations centred at 1420.4058 MHz is 1.4 MHz. The center volocity was set at $\it{v}_{c}(l)$ = +80 - (1.6$\times l $) for each longitude. The spectra has velocity width of 341 km s$^{-1}$. Each single observation of a survey field has integration time 200 seconds.

\begin{figure}
\vspace{80mm}
\begin{picture}(80,80)
\put(-25,220){\includegraphics{f1a-bw.eps}}
\put(230,150){\includegraphics{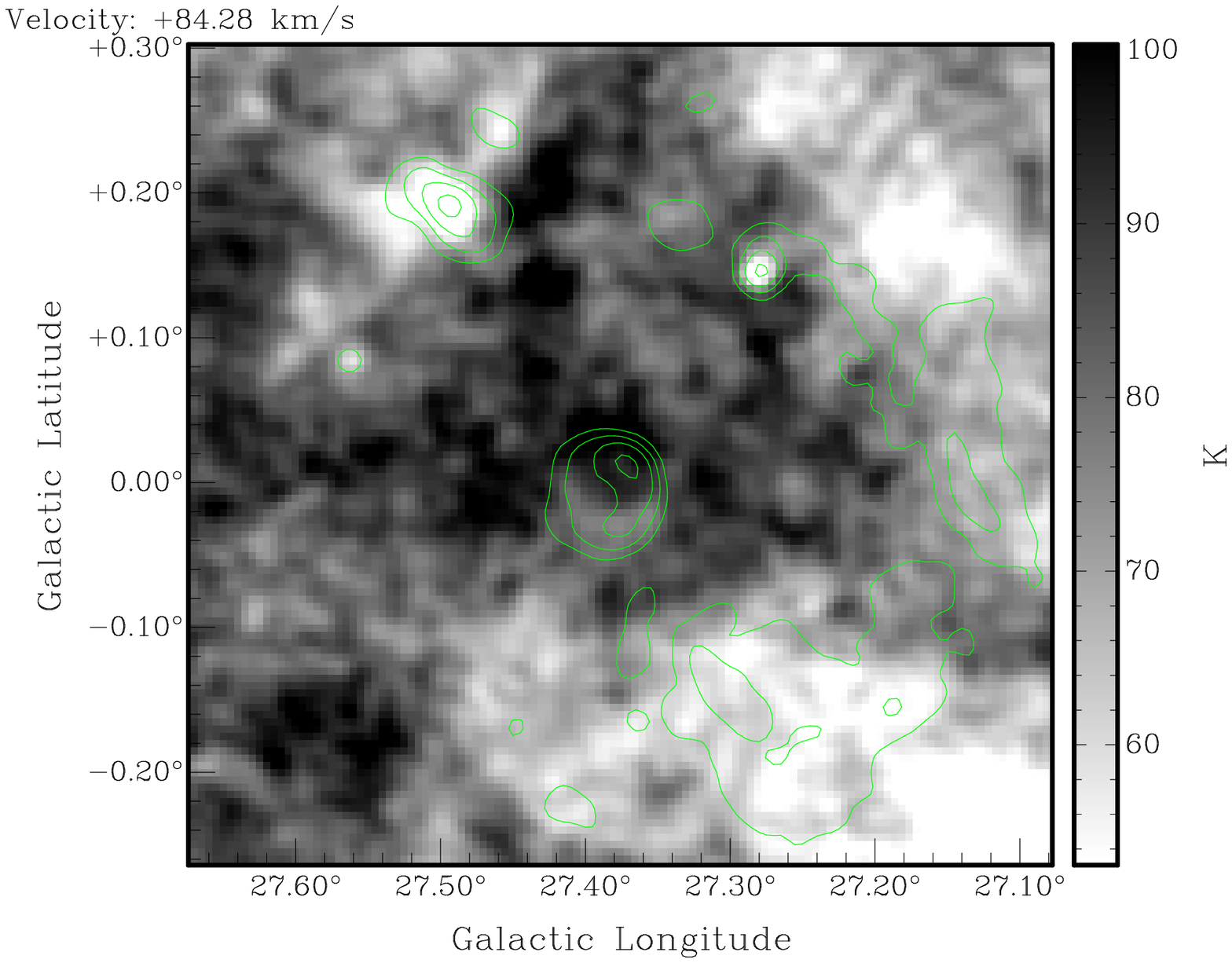}}
\put(-30,-70){\includegraphics{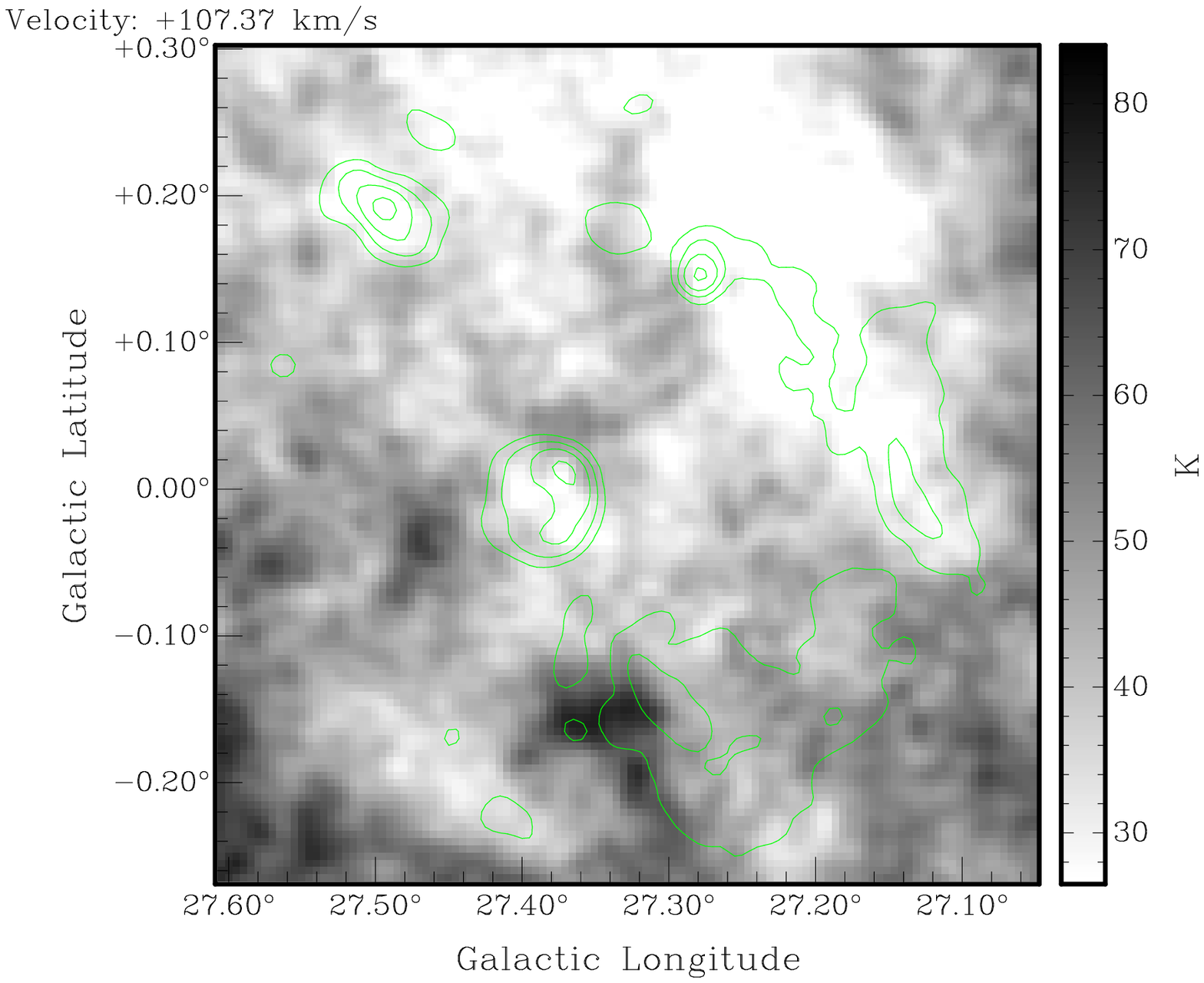}}
\put(230,-70){\includegraphics{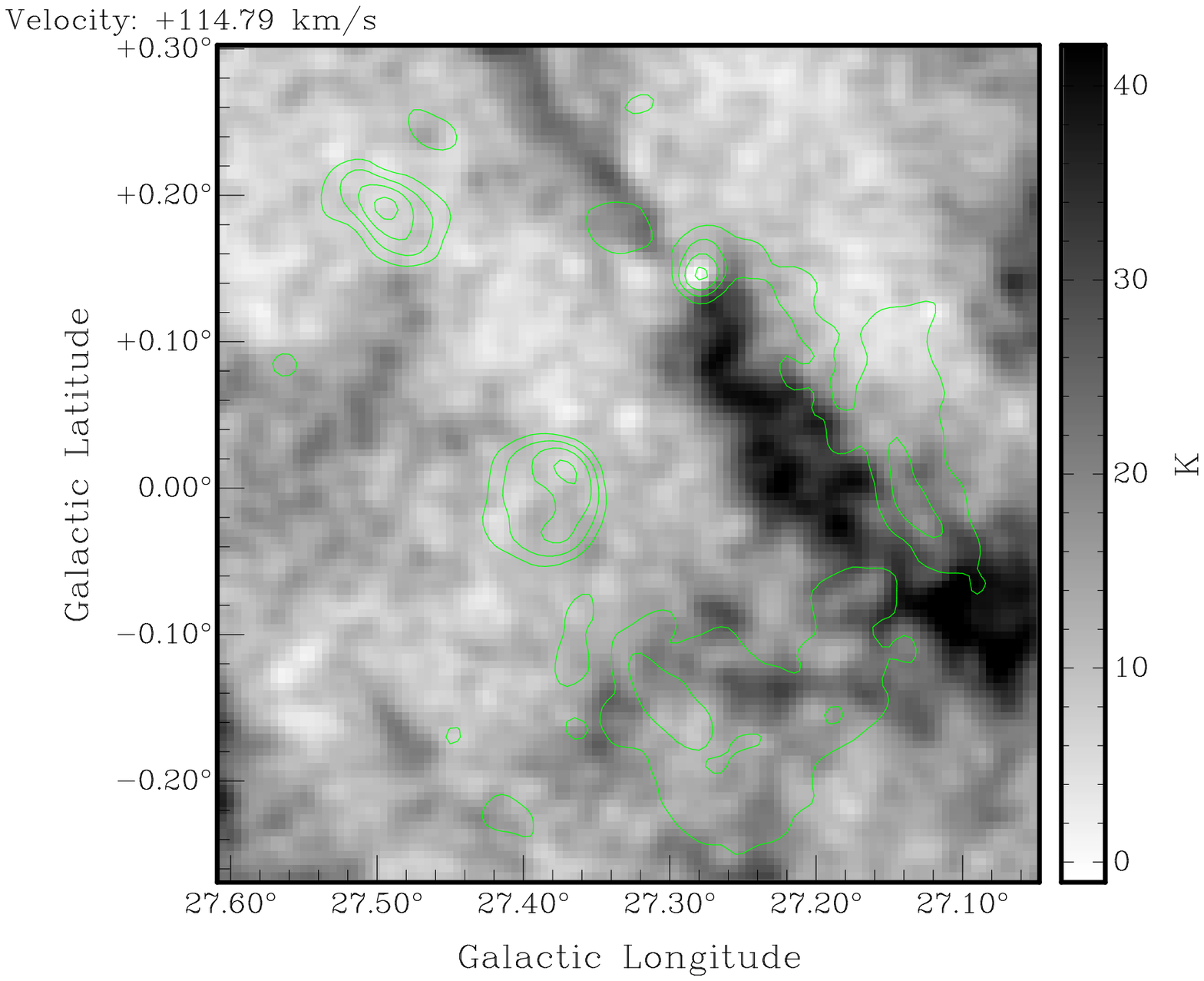}}
\end{picture}
\caption{The 1420 MHz continuum image (upper left) of Kes 73 and the images of HI emission at 
84, 107 and 114 km/s respectively. The HI maps have superimposed contours (28, 40, 60, 100 K)
of the 1420 MHz continuum emission to show the SNR and HII regions. The solid and dashed lines in the first panel show the source and background areas for the HI spectra from boxes 1, 2 and Kes 73.}
\end{figure}

\begin{figure}
\vspace{120mm}
\begin{picture}(80,80)
\put(-30,320){\includegraphics{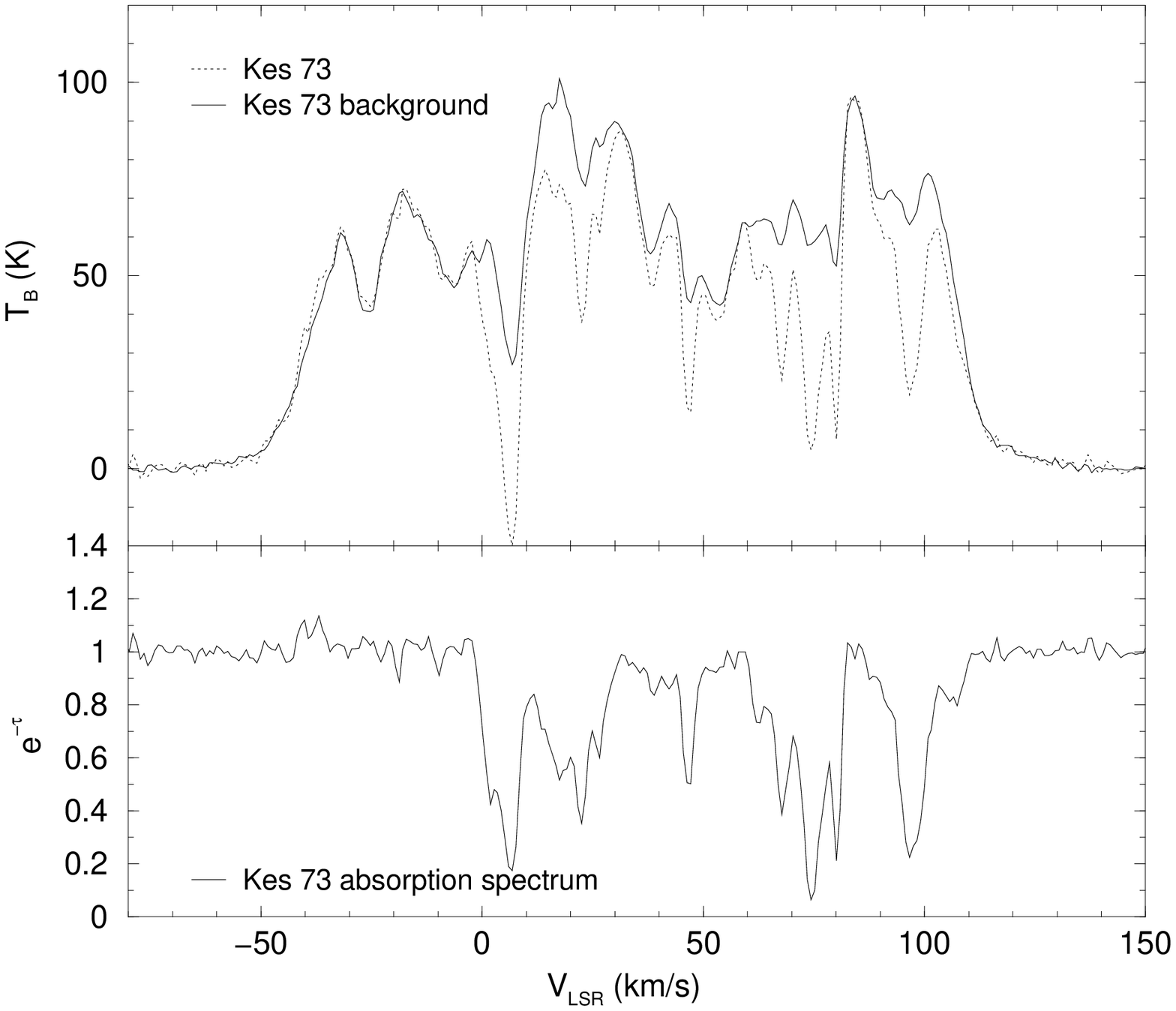}}
\put(240,490){\includegraphics{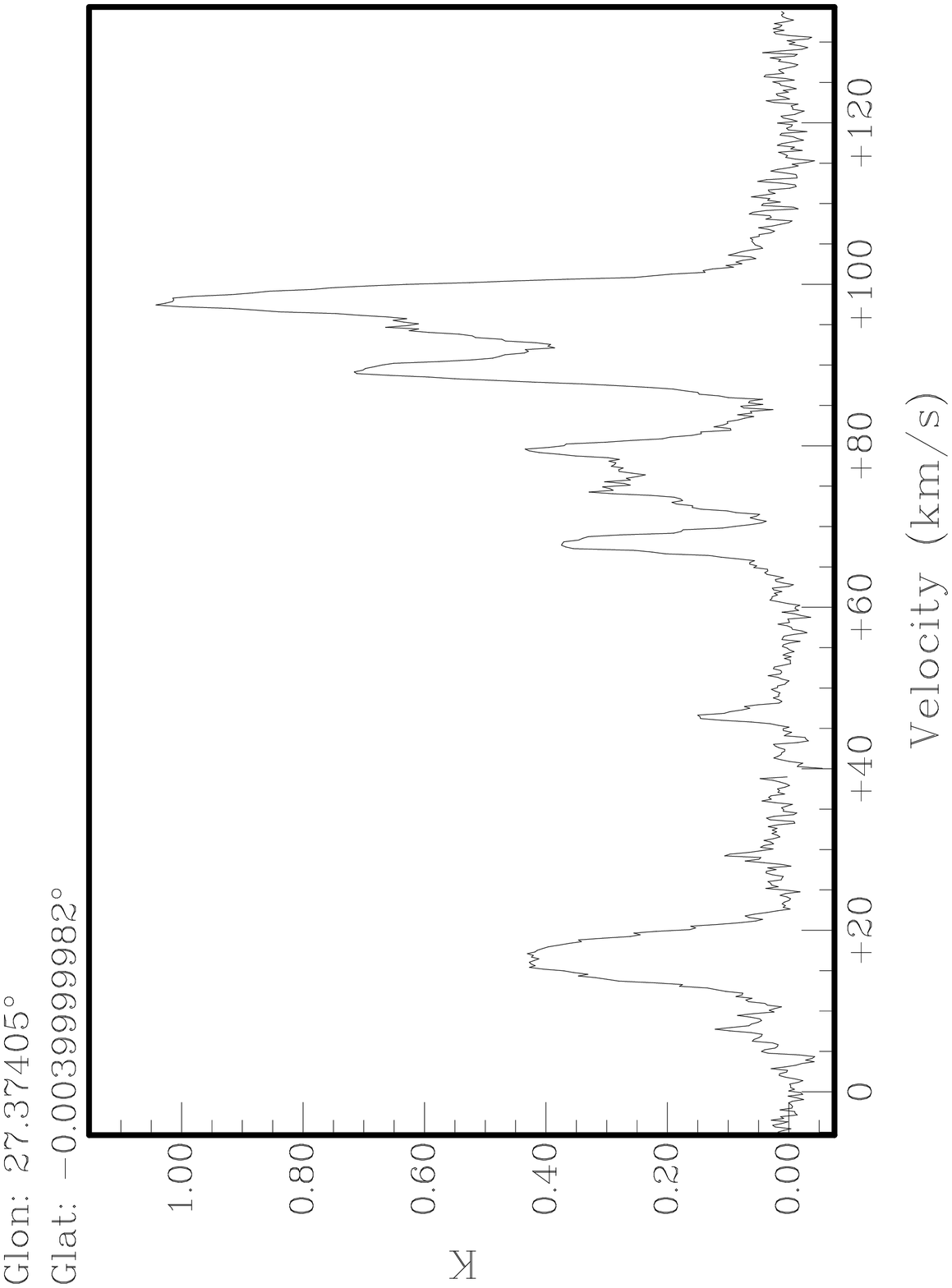}}
\put(-30,170){\includegraphics{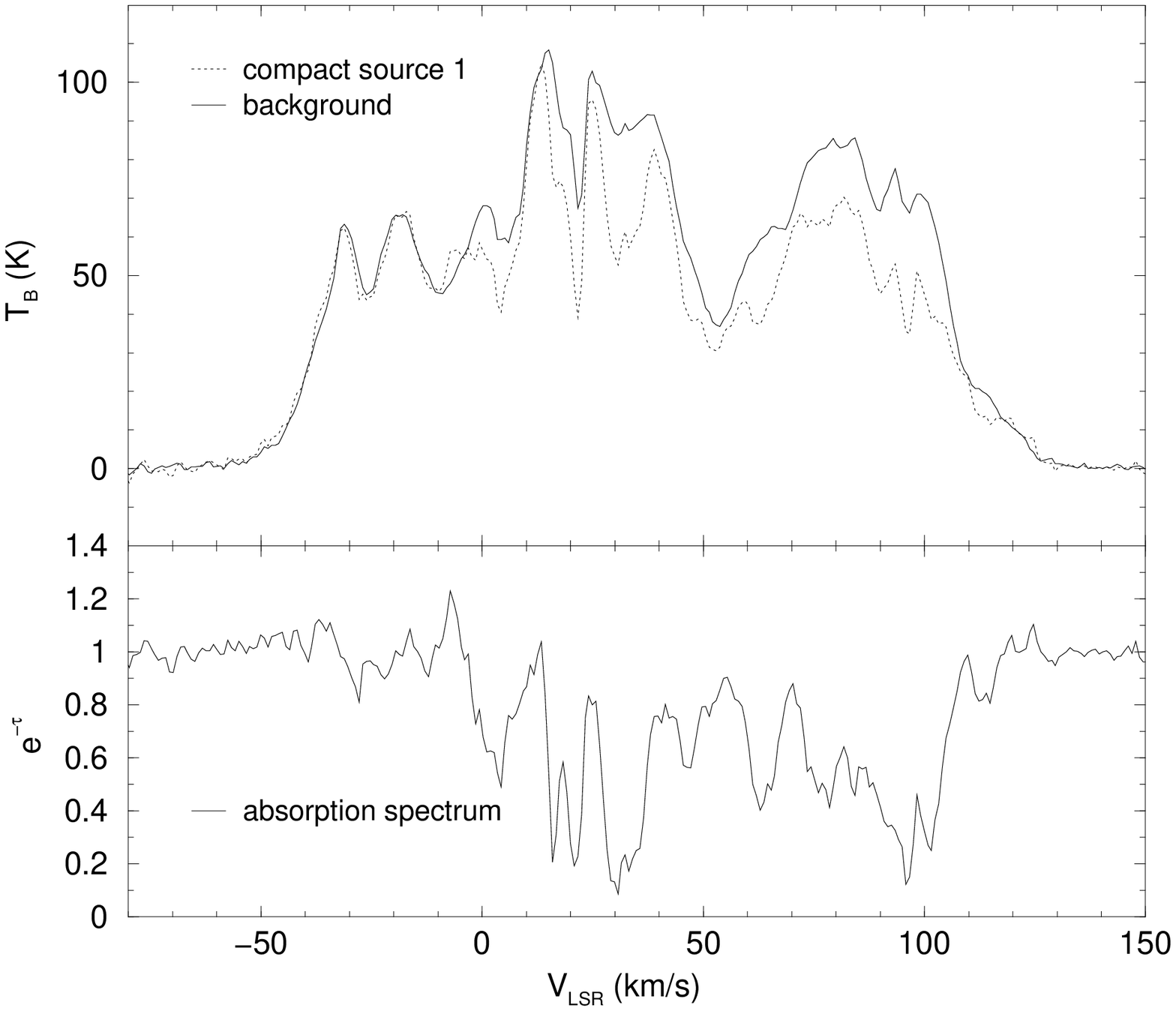}}
\put(240,170){\includegraphics{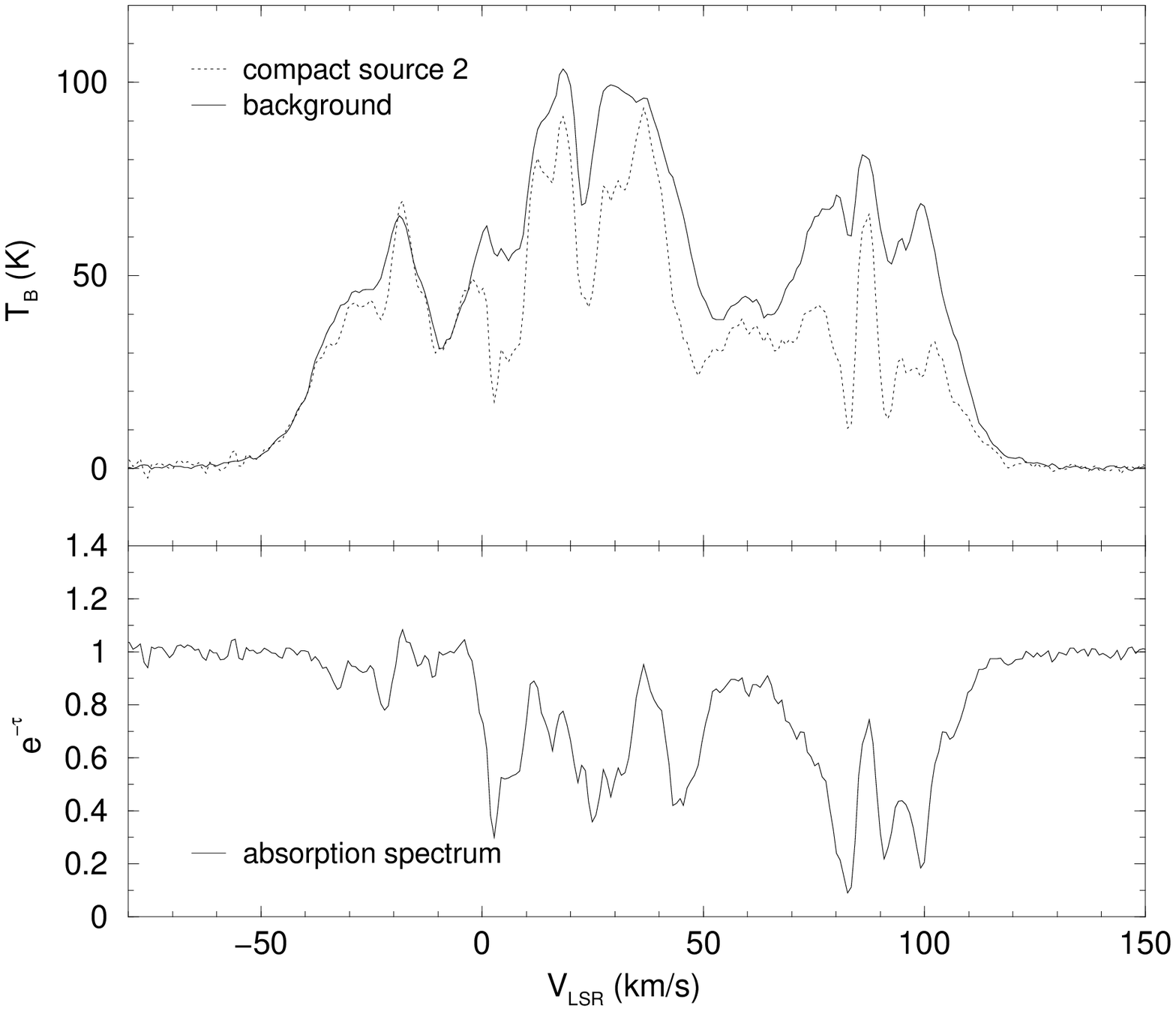}}
\put(-30,-10){\includegraphics{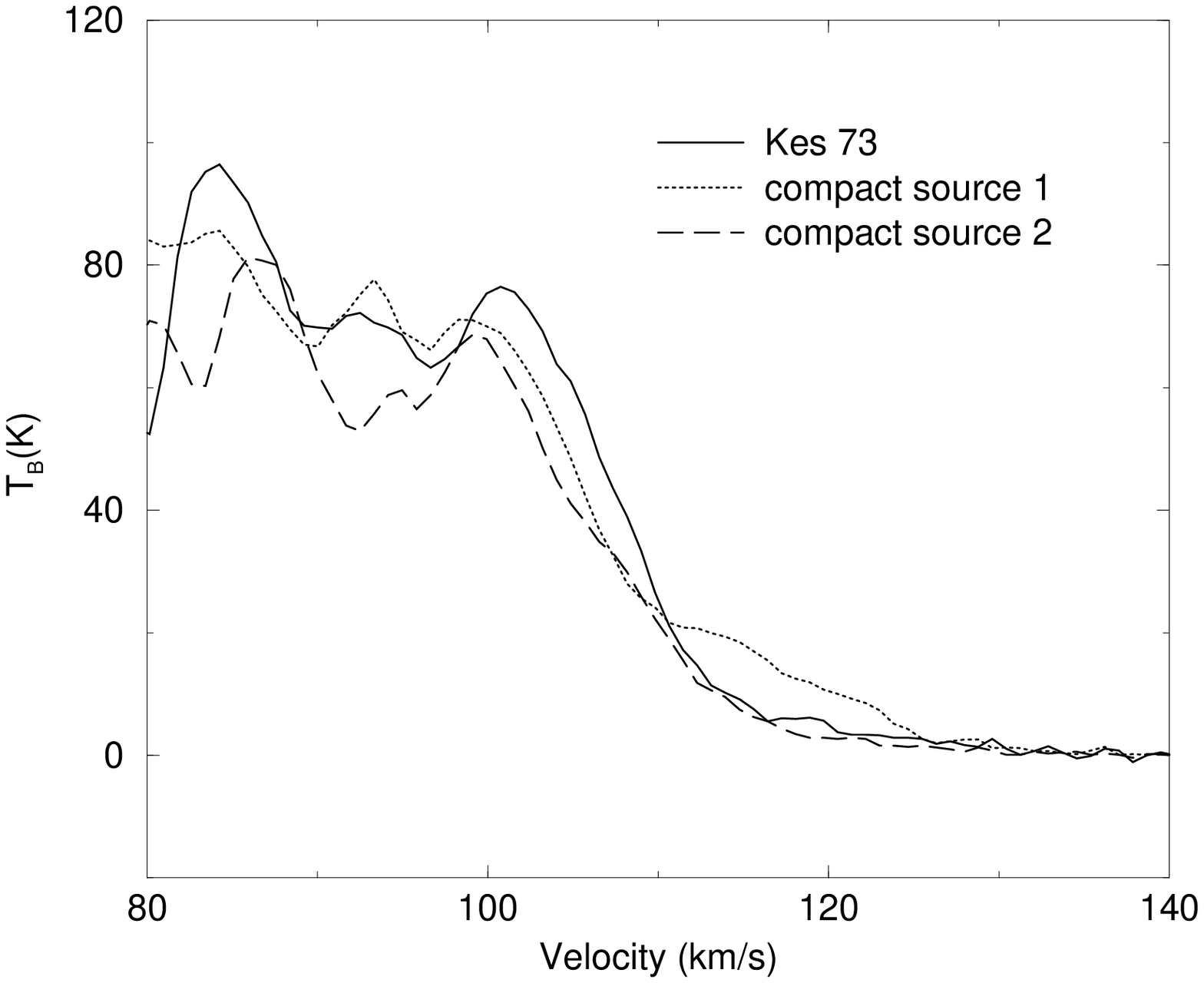}}
\put(235,-10){\includegraphics{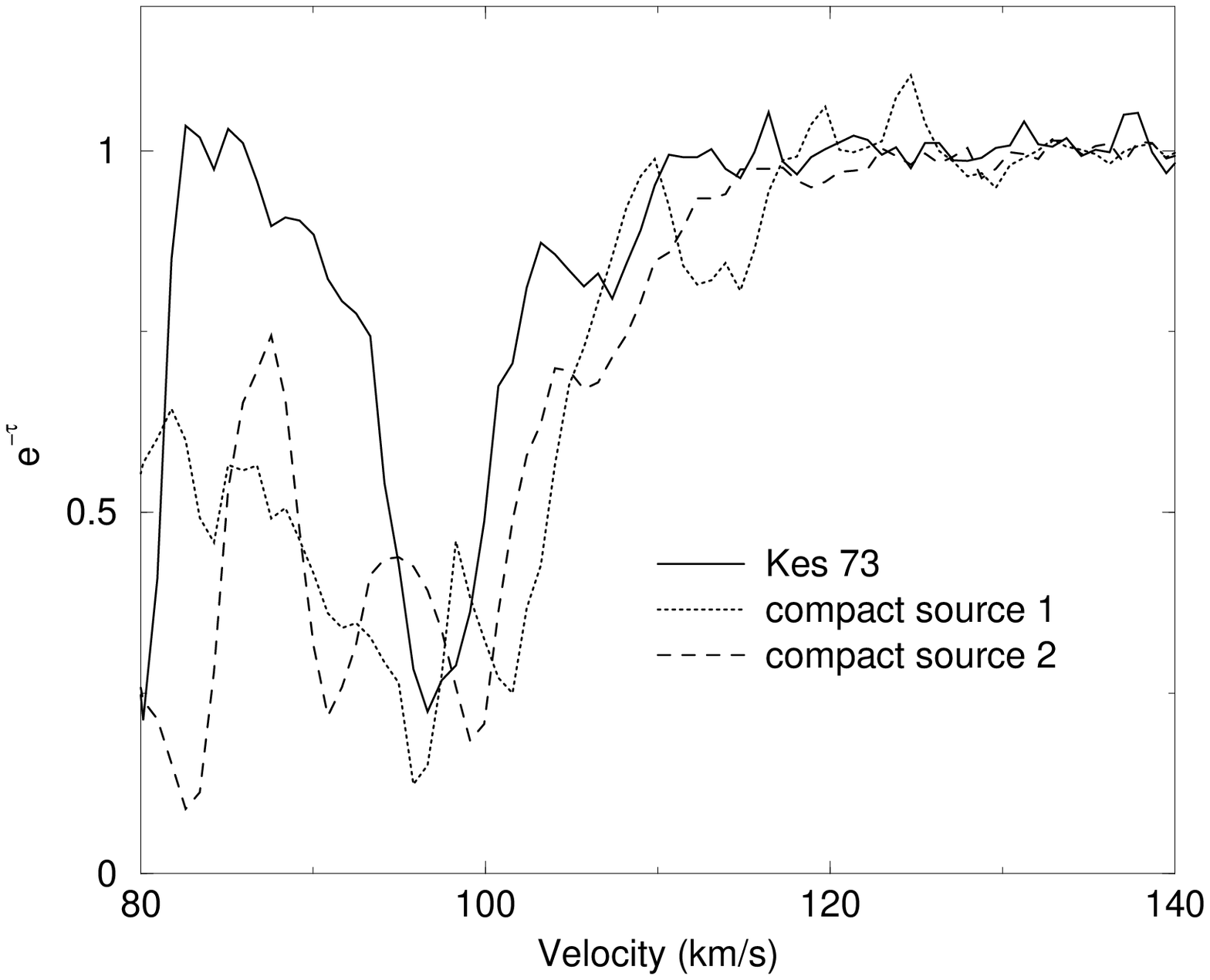}}
\end{picture}
\caption{HI spectra of: Kes 73 (top left); G27.276+0.148 (compact source 1, middle left); and 
G27.491+0.189 (source 2, middle right). The $^{13}$CO emission spectrum of Kes 73 is shown at
 top right. The details of the emission and absorption spectra of the three regions are shown
at bottom left (panel 5) and bottom right (panel 6), respectively.}
\end{figure}

\section{Analysis and Results}
Fig 1. shows the 1420 MHz continuum image (upper left) of Kes 73 and two nearby HII regions
 (G27.276+0.148 and G27.491+0.189). It also shows 
the HI emission at 84 km/s (upper right), 107 km/s (lower left) 
and 114 km/s (lower right), respectively. The HI image at 107 km/s shows clear absorption
 features associated with Kes 73 and the two HII regions.  
At 114 km/s, the HI distribution is quite inhomogeneous: there is bright emission at the 
position of G27.276+0.148 (the depression here is due to HI absorption against G27.276+0.148);
 but at the positions of Kes 73 and G27.491+0.189 the HI emission is much fainter. 
The HI images in the whole velocity range show that absorption features associated
 with the supernova remnant Kes 73 and the two HII regions appear up to the tangent point
velocity ($\sim$ 110 km/s), 
giving that all three objects are located beyond the tangent point.
So, the lower distance limit to Kes 73 is 7.5 kpc, assuming a circular Galactic rotation
 curve model with the IAU adopted value of solar circular velocity V$_{0}$=220 km/s 
and galactocentric radius of the sun R$_{0}$=8.5 kpc. 
%most recent parameters (V$_{0}$=214$\pm$7 km/s and R$_{0}$=7.6$\pm$0.3 kpc).
 At 84 km/s, bright HI emission covers Kes 73 and the two HII regions:  there are clear HI 
absorption depressions at the positions of the two HII regions but no depression at Kes 73.
 This implies that the HI at 84 km/s is at the far kinematic distance, since Kes 73 shows 
absorption up to the tangent point velocity. It also implies that 
Kes 73 is in front of the HI at 84 km/s, and the two HII regions are behind it.     

We confirm the above conclusions using the HI emission and absorption spectra and the
 $^{13}$CO emission spectra of Kes 73, G27.276+0.148 and G27.491+0.189. %The methods to construction HI spectra have been given in detail in Tian et al (2007b) and Leahy \& Tian (2007).
The full velocity range absorption spectra are shown in panels 1, 3 and 4 of Fig. 2. 
%The three absorption spectra show no clear absorption features at negative velocities, supporting evidence that all three objects are located inside the solar circle.

The errors in these spectra are systematic and not statistical, so are best estimated 
by the fluctuations
in the spectra at velocities where there is no real emission or absorption. For the emission
spectra, the errors per channel are about 2 K. For the absorption spectra, the errors in $e^{-\tau}$
depend on the strength of the continuum emission. Here the errors in  $e^{-\tau}$ are $\sim$0.08 for
Kes 73, $\sim$0.1 for compact source 1 and  $\sim$0.04 for compact source 2. 

Details of the emission and absorption spectra of the three regions are 
shown by panels 5 and 6 of Fig. 2 for the velocity range 80 to 140 km/s. 
Panel 5 shows that the HI emission in the directions 
of Kes 73 and compact source 2 decreases the same way, in the velocity range 104 to 115
km/s. Given a velocity dispersion for the HI gas of $\sim$5 km/s, this yields a
tangent point velocity of $\sim$110 km/s.  Compact source 1 follows the same trend but
has in addition an extra ($\sim$10 km/s wide) HI emission component centered on a
velocity of $\sim$115 km/s.
Panel 6 of Fig. 2 shows the three HI absorption spectra. These show absorption up to 
the tangent point velocity for all three directions, showing 
all three objects are beyond the tangent point. We note that the absorption at 115 km/s
in compact object 1 is consistent with the extra HI emission near that velocity (seen
in panel 5), which is not related to the tangent point. Panel 4 of Fig. 1 clearly shows 
an extended HI filament causing both this extra emission and the extra absorption feature
for compact source 1. 

Prominent HI emission at $\sim$84 km/s with no associated absorption is present in the
Kes 73 HI spectra. This limits the distance of Kes 73 between the far distance for 84 km/s 
and the distance of the tangent point. 
HI emission in the same velocity range (80-90 km/s) appears in the emission spectra of
both HII regions. In this velocity range there is obvious absorption, showing that both
HII regions are behind this gas. 
We also find that a CO cloud (at 89$\pm$2 km/s) in the direction of Kes 73 is behind Kes 73,
since it produces no respective HI absorption. 
This cloud shows absorption in the spectra of the HII regions, consistent with the conclusions
from the HI spectra that the HII regions are behind the gas in this velocity range. 
The recombination-line velocities of the two HII regions have been obtained previously 
(Lockman 1989). Our results show that the two HII regions are located at the far kinematic
 distance of their recombination-line velocities (i.e. 36.1$\pm$0.4 km/s for G27.276+0.148 
and 34.0$\pm$0.8 km/s for G27.491-0.189).

In summary, the HI and CO spectra provide strong support to Kes 73 being located beyond 
the tangent point in the velocity range $>$89 km/s. In the circular rotation model with
V(R)= 220 km/s for all galactocentric distances R and $R_0$=8.5 kpc, this gives 
upper and lower distance limits 
for Kes 73 of 9.8 kpc and 7.5 kpc.

%By comparison of the HI absorption spectrum with the CO emission spectrum, we find all major CO clouds have respective HI absorption in the direction of the two bright HII regions G27.276+0.148 and G27.491-0.189.

\section{Discussion}  

The distance to Kes 73 in the galactic circular rotation model is 7.5 to 9.8 kpc, revised 
upward from the values of 6 to 7.5 kpc also based on the same rotation model
(Sanbonmatsu \& Helfand 1992). However
the Galaxy is likely to have non-circular motions due to the presence of a galactic bar 
and due to spiral arm velocity perturbations. 
The observed velocity field of the outer galaxy has been
studied by Brand \& Blitz 1993, but this does not cover the inner galaxy region of interest
for Kes 73. Thus one must rely on methods such as modeling the l-V diagram using 
a galactic gravitational potential, as done in Weiner \& Sellwood (1999). 
Fig. 8 of Weiner \& Sellwood (1999) shows their resulting radial 
velocity distribution in the inner 8 kpc by 8 kpc region of the Galaxy. Using our
limits for Kes 73 ($>$89 km/s and on the far side of the
tangent point), we find distance limits of 7.1 kpc$<$d$<$8.1 kpc. These are reduced from
the circular rotation model values since the velocity contours in the Weiner \& Sellwood
model are shifted toward the Sun in this region by 0.4 to 1.5 kpc.

However, the Weiner \& Sellwood model predicts a tangent point velocity of 100 km/s in the direction
of Kes 73 (l=27.4$^\circ$), whereas our HI spectra clearly show emission and absorption 
out to at least 110 km/s. 
Using a flat circular rotation curve model with constant $V(R)$=220 km/s
yields a tangent point velocity of 119 km/s at l=27.4$^\circ$. If we use a more complex
circular $V(R)$ for $R<R_0$, such as the one in Weiner \& Sellwood (1999) with
V(R$_0$)=220 km/s and V(R$\sim$4kpc)=200 km/s, we obtain a tangent point velocity of 
99 km/s. The observed HI spectra indicate a tangent point velocity
of $\simeq$110 km/s, which is obtained with circular rotation and V(R$\sim$4kpc)= 210 km/s.
In this last case, which is the most consistent with our observations,
the distance to the point at 89 km/s is 9.4 kpc, so that Kes 73 is located between
7.5 and 9.4 kpc. 

Next we discuss implications of the revised distance, which we write as $d_{8.5}$=d/8.5 kpc.
Kes 73 is nearly circular in X-rays with a diameter of 4.5 arcmin, which is best determined 
from the Chandra image of Kes 73 (Morii et al. 2003). 
This is consistent with the determination using ROSAT High Resolution Imager and 
radio images (Helfand et al. 1994).
Kes 73 is likely to be the result of a type II SNe, with progenitor massive enough to 
produce a magnetar.
Expansion velocities for type II SNe range from $\sim$5000 to 10000 km/s. These serve as
upper limits to the actual expansion velocity of Kes 73 and yield the lower limits to age
of 1100$d_{8.5}$ yr and 540$d_{8.5}$ yr, respectively.
The mass of swept up interstellar medium (ISM) for Kes 73 
is $M_{sw}=23 d_{8.5}^3 n M_{\odot}$, with $n$, the ISM density
in units of cm$^{-3}$, so if Kes 73 has an ejecta mass $M_{ej}$ of a few $M_{\odot}$, the remnant
should be well into the transition to the adiabatic expansion phase.
The Sedov model gives lower limits to the shock velocity and upper limits to the
age. Using a Sedov model to reproduce the observed radius, with explosion energy $E$ and
parameter $\epsilon=E/(10^{51} \rm{erg}) \times 1/n$, yields age 1100
$d_{8.5}^{2.5} \epsilon^{-0.5}$ yr. This also
predicts a shock temperature of 4.4$d_{8.5}^{-3} \epsilon$ keV and shock velocity of 
2100 $d_{8.5}^{-1.5} \epsilon^{0.5}$km/s.  The measured
X-ray spectrum of Kes 73 (Gotthelf \& Vasisht 1999) has an electron
temperature of $\sim$0.7 keV, but this is likely an underestimate of the shock (ion) 
temperature due to lack of electron-ion equilibration. 
The alternate of a shock temperature of 0.7 keV can be achieved, 
if we adjust $\epsilon$ to 0.16, i.e. either reduced
explosion energy or increased ISM density by a factor of 6. For this case, this
increases the age to 2600 yr. 

In summary, for standard explosion energy and ISM density $n$=1 ($\epsilon=1$),
the upper (Sedov) age limit is 1100 yr and the lower (free expansion) age limit
is 500-1000 yr. Instead, if electron-ion equilibration is assumed the upper age limit
increases to 2600 yr and the explosion energy divided by ISM density is very low 
($\epsilon=0.16$). The emission lines detected in Kes 73 argue for ejecta dominated emission
(Gotthelf \& Vasisht 1997), so that $M_{sw}<M_{ej}$. This implies n$<$1 so that for a normal 
explosion energy, the parameter $\epsilon$ is also larger than 1. 
This gives an upper age limit of 
$\sim$1100 yr, so that the actual age of Kes 73 is likely 500-1000 yr.  

The main results here are larger distance to Kes 73 and the younger age than those previously
inferred. The young age (500-1000 yr) raises the possibility that Kes 73 was observable
as a historical supernova between 1000 AD and 1500 AD. If it had a typical maximum-light
absolute V magnitude for type II SNe of -17 to -18, its maximum-light V magnitude at 
8.5 kpc would be $\sim -3$ uncorrected for extinction. The visual extinction derived from the
X-ray column density of $2-3\times10^{22}$cm$^{-2}$ (Gotthelf \& Vasisht 1997)
is $A_V \sim 12$ magnitudes, resulting in a maximum light
V magnitude of $\sim$9, so that this SN would not have been noticed during the time period it 
likely exploded.
The larger distance implies a larger X-ray luminosity for AXP 1E 1841-045 than
previously quoted, by a factor of $(8.5/7)^2 \sim 1.5$. This implies a somewhat
larger magnetic field decay rate in the magnetar model or larger accretion rate
in accretion based models than for the luminosity based on the previous smaller distance
for Kes 73/ AXP 1E 1841-045.
The tighter constraint on age of Kes 73 (500-1000 yr) further 
supports the arguments of Gotthelf et al (1999) that AXP 1E 1841-045 is the youngest of
the currently known AXPs and soft gamma-ray repeaters (SGRs), and also raises the question of
whether AXP 1E 1841-045 is in a pre-SGR phase, so that we might expect a very
active SGR phase from this object in the next several 1000 years.

\begin{acknowledgements}
DAL and WWT acknowledge support from the Natural Sciences and Engineering Research Council of Canada. WWT thanks the Natural Science Foundation of China for support.  
This publication makes use of molecular line data from the Boston University-FCRAO Galactic Ring Survey (GRS). 
The GRS is a joint project of Boston University and Five College Radio Astronomy Observatory.  
The NRAO is a facility of the National Science Foundation operated under cooperative agreement by Associated Universities, Inc. 
\end{acknowledgements}

\end{document}